\documentclass[11pt]{article}
\usepackage[left=2cm, right=2cm, bottom=2cm ,top = 2cm, footskip=1cm]{geometry}
\usepackage[font=small,labelfont=bf]{caption}
\usepackage[table]{xcolor} 
\usepackage{setspace}
\usepackage{chngcntr}
\usepackage{array,multirow}
\usepackage[hidelinks]{hyperref}

\usepackage{enumitem}
\usepackage{graphicx}
\usepackage{amsmath}
\usepackage{titlesec}
\titleformat{\paragraph}
{\normalfont\normalsize\bfseries}{\theparagraph}{1em}{}
\titlespacing*{\paragraph}
{0pt}{3.25ex plus 1ex minus .2ex}{1.5ex plus .2ex}

\usepackage[super,sort&compress]{natbib}
\setcitestyle{open={},close={}}

\usepackage[section]{placeins}

\title{Comparing Methodological Variations in Seizure Onset\\Localisation Algorithms using intracranial EEG}
\author{Sarah J. Gascoigne$^{1*}$, Manel Vila-Vidal$^{2,3}$, Nathan Evans$^1$,\\ 
Christopher Thornton$^1$, Heather Woodhouse$^1$, Billy Smith$^1$,\\
Anderson Brito Da Silva$^4$, Rhys H. Thomas$^4$, Kevin Wilson$^5$,\\
Peter N. Taylor$^{1}$, Adrià Tauste Campo$^{3}$, Yujiang Wang$^1**$}
\date{}

\begin{document}

\maketitle

\begin{enumerate}
\item{CNNP Lab (www.cnnp-lab.com), School of Computing, Newcastle University, Newcastle upon Tyne, United Kingdom}
\item{BrainFocus Labs, S.L. (www.brainfocus.tech), 08750, Barcelona, Spain}
\item{Computational Biology and Complex Systems, Department of Physics, Universitat Politècnica de Catalunya, 08028, Barcelona, Spain}
\item{Faculty of Medical Sciences, Newcastle University, United Kingdom}
\item{School of Mathematics, Statistics \& Physics, Newcastle University, United Kingdom}
\end{enumerate}

\begin{itemize}[label={}]
\item $^*$ Sarah Jane Gascoigne (ORCID ID 0000-0003-1013-1875)\\
Urban Sciences Building, 
1 Science Square Newcastle upon Tyne, NE4 5TG\\
Email: S.Gascoigne@newcastle.ac.uk
\item $^{**}$ Yujiang Wang (ORCID ID: 0000-0002-4847-6273)\\
Urban Sciences Building, 
1 Science Square Newcastle upon Tyne, NE4 5TG\\
Tel: (+44) 191 208 4141  \hskip5em
Email: Yujiang.Wang@newcastle.ac.uk
\end{itemize}
\begin{center}
manel.vilavidal@brainfocus.tech; N.Evans5$^a$; Chris.Thornton$^a$; H.Woodhouse$^a$; B.Smith16$^a$, Anderson.Brito-da-Silva$^a$; Rhys.Thomas$^a$; Kevin.Wilson$^a$; Peter.Taylor$^a$; adria.tauste@gmail.com;

$^a$ @newcastle.ac.uk
\end{center}
\noindent{
We confirm that we have read the Journal’s position on issues involved in ethical publication and affirm that this report is consistent with those guidelines.\\
None of the authors has any conflict of interest to disclose.}
\newpage

\begin{doublespace}
\section*{Abstract}
During clinical treatment for epilepsy, the area of the brain thought to be responsible for  pathological activity - known as the seizure onset zone - is identified. This identification is typically performed through visual assessment of EEG recordings; however, this is time consuming and prone to subjective inconsistency. Automated onset localisation algorithms provide objective identification of the onset location by highlighting changes in signal features associated with seizure onset. In this work we investigate how methodological differences in such algorithms can result in different onset locations being identified.

We analysed ictal intracranial EEG (icEEG) recordings in 16 subjects (100 seizures) with drug-resistant epilepsy from the SWEZ-ETHZ public database. Through our analysis, we identified a series of key methodological differences that must be considered when designing or selecting an onset localisation algorithm. These differences were demonstrated using three distinct algorithms that capture different, but complementary, seizure onset features: Imprint, Epileptogenicity Index (EI), and Low Entropy Map (LEM). We assessed methodological differences (or Decision Points), and their impact on the identified onset locations.

Our independent application of all three algorithms to the same ictal icEEG dataset revealed low agreement between them: 27-60\% of onset channels showed minimal or no overlap. Therefore, we investigated the effect of three key methodological differences: (i) how to define a baseline, (ii) whether low-frequency components are considered, and finally (iii) whether electrodecrement/EEG suppression at onset is considered. Changes at each Decision Point were found to substantially influence resultant onset channels ($r>0.3$). 

Our results demonstrate how seemingly small methodological changes can result in large differences in onset locations. We propose that key Decision Points must be considered when using or designing an onset localisation algorithm in the future.

\newpage 

\subsection*{Key Words}
Automatic localisation of seizure onset; Algorithmic onset localisation; Objective comparisons

\subsection*{Key Points} 
\begin{itemize}
    \item When comparing automatic onset localisation algorithms, overlap between onset locations tended to be low (1.8-2.9\% of seizures had the same/similar onset locations between pairs of algorithms).
    \item Differences between onset locations can be attributed to various methodological differences between onset localisation algorithms.
    \item Before developing or using an onset localisation algorithm, the implications of the methods underpinning the algorithm must be understood. 
\end{itemize}

\subsection*{Abbreviated Summary}
Gascoigne et al. quantitatively compare onset locations across three independent onset localisation algorithms. Each algorithm was applied to intracranial EEG recordings (16 subjects, 100 seizures). Independent analyses for each Decision Point captured the impact of the difference on resultant onset channels. 

\newpage
\section*{Introduction}
Using intracranial electroencephalography (icEEG) to identify where seizures start is a cornerstone of epileptology. It allows clinicians to detect the distinct electrophysiological features of seizure onset, providing targets for treatments such as resective surgery and supporting our understanding of the epilepsy etiology. The current gold standard for identifying the channels involved in seizure onset (onset channels) involves expert epileptologists performing a visual assessment of each recording channel of the icEEG \cite{Bartolomei2008EpileptogenicityEEG, Weiss2015SeizureStudy,desai2023expert, haut2002interrater}. In the last two decades, automated methods (onset localisation algorithms) have been developed \cite{Bartolomei2008EpileptogenicityEEG, gascoigne2024incomplete, Vila-Vidal2017DetectionIdentification, vila2020low, Weiss2015SeizureStudy, andrzejak2012nonrandomness,david2011imaging,gnatkovsky2011identification}, but are not yet widely used clinically. 

Methodological differences between onset localisation algorithms stem from different approaches and aims. Some algorithms were created to capture specific types of signal changes associated with seizure onset. For example, the Epileptogenicity Index (EI) \cite{Bartolomei2008EpileptogenicityEEG} was created to capture rapid increases in high-frequency EEG activity coupled with decreases in low-frequency activity occurring early in the seizure. However, there is variation in EEG signal behaviour at seizure onset both across- and within-individuals \cite{pelliccia2013ictal, salami2020seizure, Donos2018, velascol2000functional, perucca2014intracranial, lee2000intracranial}. Therefore, other algorithms have been created to capture more general changes using a broader range of features (e.g., \cite{gascoigne2024incomplete, Vila-Vidal2017DetectionIdentification, vila2020low}).  To date, the methodological differences between onset localisation algorithms have not been systematically investigated.


To this end, we assess three onset localisation algorithms: Imprint \cite{gascoigne2024incomplete}, Low Entropy Map \cite{vila2020low}, and Epileptogenicity Index \cite{Bartolomei2008EpileptogenicityEEG}. Using a sample dataset, we determine the extent to which they agree with respect to seizure onset locations,  identify key methodological differences between them (which we term Decision Points), and demonstrate the implications of these decisions for the resulting onset locations. Specifically, we investigate differences in the selection of baselines and how ictal activity is captured. 

We present these Decision Points as important considerations when using or designing an onset localisation algorithm.

\newpage
\section*{Methods}
\subsection*{Terminology}
For clarity, we use the following terminology throughout this work:
\begin{itemize}
    \item \textbf{onset location}: the location of onset in space as a broad term, either identified by an algorithm or otherwise.
     \item \textbf{onset channel}: icEEG channels algorithmically labelled as onset by specific algorithms. E.g. channels labelled by LEM are referred to as LEM onset channels.
    \item \textbf{onset localisation algorithm/method}: automatic method applied to icEEG timeseries to label icEEG channels as onset
\end{itemize}

\subsection*{Subjects and recordings}
We analysed ictal icEEG recordings of 100 seizures across 16 individuals with drug-resistant epilepsy from the  publicly available, short-term SWEZ-ETHZ database \cite{burrello2019laelaps} (accessible at \url{http://ieeg-swez.ethz.ch/}). Ictal icEEG recordings were available together with a short preictal period.

icEEG data were preprocessed prior to uploading to the database, this included band pass filtering between 0.5 and 150 ~Hz using a fourth order zero phase-shift Butterworth filter and downsampling (where appropriate) to 512~Hz. A common-median referencing was applied. Channels with artefacts throughout the recording were excluded from all seizures. We performed no additional preprocessing and used the data as provided. 

\subsection*{Onset localisation algorithms}
To investigate how methodological differences in algorithms can lead to differences in onset locations, we compared seizure-specific onset locations using three algorithms: Imprint \cite{gascoigne2024incomplete}, Low Entropy Map (LEM) \cite{vila2020low}, and Epileptogenicity Index (EI) \cite{Bartolomei2008EpileptogenicityEEG}. Two algorithms (Imprint and LEM) were designed by the authors of this work, the third (EI) is a well-established algorithm in this field. 
Parameters, decisions and thresholds within these algorithms were optimised using separate data independent of the 100 seizures used in this study (see Suppl. \ref{suppl:thresholds_methods}).

There are fundamental differences between these algorithms, potentially leading to non-concordant onset channel identification across algorithms. To minimize confounding factors, we first harmonized several preprocessing steps and definitions that varied among the algorithms (see Suppl. \ref{suppl:harmon}). Importantly, these harmonisations did not alter the fundamental principles of the algorithms, which are described below. 

\subsubsection*{Imprint}
The Imprint algorithm, as described in \cite{gascoigne2024incomplete}, detects changes in icEEG activity across eight EEG features (line length, energy (relative band power across bands), and $\delta$: 1-4~Hz , $\theta$: 4-8~Hz, $\alpha$: 8-13~Hz, $\beta$: 13-30~Hz, low-$\gamma$: 30-60~Hz, and high-$\gamma$: 60-100~Hz band powers). Baseline activity is estimated using 110 seconds of preictal activity for each recording channel. Brief abnormal activity in the preictal segment is automatically identified and removed from the baseline. 

Potential ictal activity is identified as deviations from the baseline using Mahalanobis distances and Median Absolute Deviation (MAD) scores (see \cite{gascoigne2024incomplete}). Ictal activity is required to persist for 80\% of a nine-second window, meaning that artefacts or spikes preceding seizure onset will not be captured as a false seizure onset. Finally, a map of ictal activity is generated across time and recording channels, from which the first channel(s) with activity are labelled as onset.

\subsubsection*{Low Entropy Map (LEM)} 
The Low Entropy Map (LEM) algorithm, as described in \cite{vila2020low}, identifies the spectro-temporal signatures of ictal onset patterns by targeting increases in signal power that are confined to a few electrode contacts. This is achieved by exhaustively inspecting different frequency bands spanning the whole spectrum and time windows starting at the clinically labelled onset time \cite{vila2020low}. Signals are first split in nine different frequency components using the following cutting-points: 0.9, 4.4, 7.4, 12.5, 27.4, 46.4, 60.3, 78.4, 101.9, and 132.5~Hz. Preictal power values across all channels and time points (starting 115 seconds before clinically labelled onset time) are pooled to create a global baseline. Then, all ictal power values are MAD-scored against the preictal baseline for each frequency band independently, thus generating a seizure activation map at each frequency band. Average MAD values are obtained in a number of nested time windows with increasing upper bounds, up to a maximum bound of 30 seconds. The time windows use a finer granularity in the vicinity of seizure onset to capture short-lasting fast activity.

The global activation (GA) and the activation entropy (AE) are used to characterise the power increase and the spatial spread of activations for each time-frequency window, respectively. These two metrics are used to automatically identify the ictal temporal and spectral features. Accumulation across time-frequency windows is used to delineate the onset location (see \cite{vila2020low}).

\subsubsection*{Epileptogenicity Index (EI)}
The Epileptogenicity Index (EI), as described in \cite{Bartolomei2008EpileptogenicityEEG},
detects changes in signal activity ratios between slow ($<12.5$~Hz) and fast ($\geq12.5$~Hz) oscillations compared to a cumulative baseline. This algorithm uses the Page-Hinkley Algorithm \cite{page1954continuous, hinkley1970inference} to identify rapid changes in activity over time.

Interestingly, EI combines the scale of the change in activity with the time between first activity and commencement of activity in each channel to create the onset. Therefore, the highlighted channels are not necessarily the first with activity; instead the onset location might comprise of channels where activity commenced slightly later but with a much larger increase in activity.

\subsection*{Descriptive analyses: size of onsets and overlap across algorithms}
\begin{figure}
    \centering
    \includegraphics[width=0.75\linewidth]{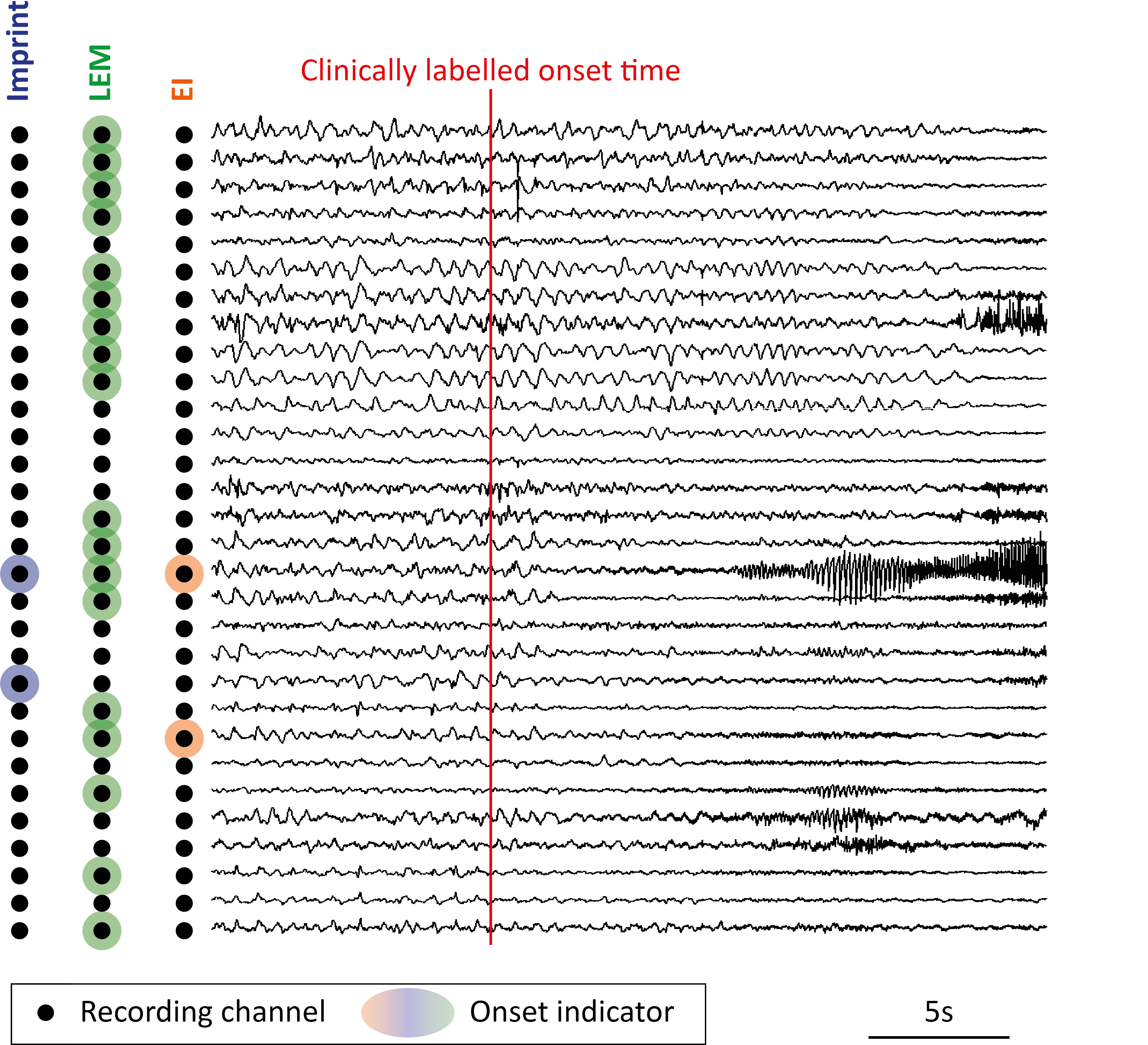}
    \caption[Example seizure with Imprint, Epileptogenicity Index (EI), and Low-Entropy Map (LEM) onset channels highlighted.]{\textbf{Example seizure with Imprint, Epileptogenicity Index (EI), and Low-Entropy Map (LEM) onset channels highlighted.} A representative subset of 30 channels is presented. Intracranial EEG timeseries (right) of an example seizure with the Imprint (blue), EI (orange), and LEM (green) onset channels highlighted (left). Differences in onsets across algorithms are seen.}
    \label{fig:Ons_comp}
\end{figure}

We first ran the three algorithms across all seizures and inspected the size of onset locations (i.e., number of onset channels). We then investigated whether onset locations were consistent across algorithms. Figure \ref{fig:Ons_comp} displays an example seizure with Imprint, EI, and LEM onset channels indicated. Per pair of algorithms, we assessed overlap between them by computing the proportion of each onset captured within the other, in both directions (see Suppl. \ref{suppl:cat_seizures}).

\subsection*{Decision Points: defining methodological differences between algorithms}

When designing an algorithm, several decisions must be made regarding the type of ictal activity to capture, the baseline to score ictal changes against, and whether to apply any exclusion criteria. For the three algorithms included in this study, we identified three key Decision Points that define each algorithm and may account for the differences in their outputs. For conciseness, we have selected a pair of algorithms to compare for each Decision Point. That is, since the decision points are YES/NO in nature, there will always be two algorithms using the same methodology. In each case, we selected one of the two as a representative example. See Table \ref{tab:dec_point} for all Decision Points considered and the pair of algorithms selected for comparison in each case. 

\begin{table}[]
    \centering
    \begin{tabular}{c|ll|ccc}
        \hline
        Decision Point & \multicolumn{2}{l|}{Decision}& Imprint & LEM & EI \\
        \hline
        1a& \multirow{2}{4em}{Baseline}& Fixed?& \cellcolor{gray!25}\textbf{Y}& Y& \cellcolor{gray!25}\textbf{N}\\
        1b& & Channel-specific?& \cellcolor{gray!25}\textbf{Y}& \cellcolor{gray!25}\textbf{N}&Y  \\
        2& \multicolumn{2}{l|}{Increases in low frequency?} & Y& \cellcolor{gray!25}\textbf{Y}& \cellcolor{gray!25}\textbf{N}  \\
        3& \multicolumn{2}{l|}{Electrodecrements?}& \cellcolor{gray!25}\textbf{Y}& \cellcolor{gray!25}\textbf{N} & Y\\
        \hline
    \end{tabular}
    
    \caption{\textbf{Table showing Decision Points assessed.} Decision point 1a: Y = Fixed segment of preictal activity used as baseline, N = Moving window of preictal activity used as baseline. Decision Point 1b: Y = Baseline is created for each channel independently (channel-specific), N = Baseline is created across all channels (global). Decision Point 2: Y = Method includes increases in low-frequency activity in activity detection, N = Method does not include increases in low-frequency activity. Decision Point 3: Y = Decreases in activity (electrodecrements) are considered in ictal detection, N = Only increases in activity are considered.
    Algorithms selected for each comparison are shown in bold. Corresponding Decision Point identifiers are listed. }
    \label{tab:dec_point}
\end{table}

\subsubsection*{Decision Point 1: Definition of baseline activity.}

Baseline activity is often used as a reference to identify ictal activity. Each of the onset localisation algorithms use a baseline, but they differ in two key aspects that can be summarised as the temporal (i.e., time windows used: fixed vs. moving baseline) and spatial (i.e., channels used: channel-specific vs. global baseline) definition of the baseline.

For each comparison, we aimed to assess whether differences in baseline selections might translate into differences in onset locations. Given that an algorithm uses a specific baseline, we hypothesised that changes from this baseline would be more pronounced in this algorithm’s onset channels, compared to another algorithm's onset channels. See Supplementary \ref{suppl:baseline_methods} for further details on how ictal activity was scored against the baseline. 

\noindent{\textbf{Decision Point 1a: Fixed vs. moving baselines}} 

First, we investigated whether defining a fixed or moving baseline impacted resultant onset channels. We compared Imprint and EI, two algorithms that use channel-specific baselines but with different time windows: Imprint applies a fixed preictal window, while EI uses a cumulative moving window (see Fig.~\ref{fig:baseline}A). To enable an easier comparison, we focussed on onset channels identified by Imprint and/or EI (see Fig.~\ref{fig:baseline}C).

As the EEG signal features used in each of these algorithms are not consistent, we required a common and consistent feature to compare the algorithms. This consistent feature was not used to alter the algorithms, instead it was used in a post-hoc analysis to best represent common features captured by both algorithms. Given that EI is more influenced by increases in high-frequency activity, we used broad-band high-frequency band power (12.5-100~Hz) as the consistent feature. 

\textbf{Post-hoc analysis:} For each seizure we obtained a fixed, channel-specific, preictal baseline (120 to 10 seconds before marked onset) from which we created a baseline distribution of broad-band high-frequency band powers (see Fig.~\ref{fig:baseline}A top). We captured ictal deviations by comparing early ictal activity (defined as the first five seconds of activity following the Imprint onset time) against this baseline using MAD scores. We divided onset channels into two groups: Imprint onset channels \textit{vs.} EI onset channels not in Imprint (Fig.~\ref{fig:baseline}C `Fixed' baseline case). Ictal deviations were summarised per group by taking the median across channels. Finally, we computed the difference in MAD scores between the channel groups (MAD(\texttt{[in Imprint]}) – MAD(\texttt{[in EI but not Imprint]})) to obtain a direct quantification of the Decision Point (fixed \textit{vs.} moving baseline) effect.

We used a one-tailed paired-sample Wilcoxon rank-sum test to determine if ictal deviations were substantially higher for Imprint onset channels ($H_0:$ \texttt{[in Imprint]} = \texttt{[in EI but not in Imprint]}, $H_1:$ \texttt{[in Imprint]} $>$ \texttt{[in EI but not in Imprint]}). Effect sizes ($r$) were calculated as $r=\frac{Z}{\sqrt{N}}$, where $Z$ is the test statistic and $N$ is the total sample size. Effect sizes were categorized by standard thresholds, where $0.1\leq r<0.3$ indicates a small effect, $0.3\leq r<0.5$ a moderate effect, and $r\geq0.5$ a large effect. 

We repeated this post-hoc analysis for EI. For each seizure we obtained a moving baseline (up to 10 seconds before EI onset, excluding time windows earlier than 10 seconds before clinically labelled onset time) from which we created a baseline distribution of broad-band high-frequency band powers (see Fig.~\ref{fig:baseline}A bottom). We captured ictal deviations by comparing early ictal activity (defined as the first five seconds of activity following the EI onset time) against this baseline using MAD scores. We divided onset channels into two groups: EI onset channels \textit{vs.} Imprint onset channels not in EI (Fig.~\ref{fig:baseline}C `Moving’ baseline case). Ictal deviations were summarised per group by taking the median across channels. Finally, we computed the difference in MAD scores between the channel groups (MAD(\texttt{[in EI]}) – MAD(\texttt{[in Imprint but not EI]})). See Suppl. \ref{suppl:fm_methods} for further details on how ictal activity was scored against the baseline.

We used a one-tailed paired-sample Wilcoxon rank-sum test to determine if ictal deviations were substantially higher for EI onset channels ($H_0:$ \texttt{[in EI]} = \texttt{[in Imprint but not in EI]}, $H_1:$ \texttt{[in EI]} $>$ \texttt{[in Imprint but not in EI]}), again to obtain a direct quantification of the Decision Point (moving \textit{vs.} fixed baseline) effect.

\noindent{\textbf{Decision Point 1b: Channel-specific vs. global baselines}}
Here, we investigated whether defining a channel-specific or global baseline impacted resultant onset channels. We compared Imprint and LEM, two algorithms that use fixed preictal time-windows but with different channels in the baseline: Imprint creates a baseline distribution for each channel individually, while LEM combines preictal activity in all channels to create a baseline (see Fig.~\ref{fig:baseline}B). To enable an easier comparison, we focussed on onset channels identified by Imprint and/or LEM (see Fig.~\ref{fig:baseline}D).

As the EEG signal features used in each of these algorithms are not consistent, we required a common and consistent feature to compare the algorithms. This consistent feature was not used to alter the algorithms, instead it was used in a post-hoc analysis to best represent common features captured by both algorithms. Given that the LEM algorithm uses activity in nine frequency bands, we used band-powers in each of the frequency bands as our consistent features, repeating the analysis for each band. 

\textbf{Post-hoc analysis:} For each seizure we obtained a fixed, channel-specific, preictal baseline (115 to 10 seconds before marked onset) for each of the frequency bands from which we created channel-specific baseline distributions (see Fig.~\ref{fig:baseline} B top). We captured ictal deviations by comparing early ictal activity (defined as the first five seconds of activity following the Imprint onset time) against this baseline using MAD scores. We divided onset channels into two groups: Imprint onset channels \textit{vs.} LEM onset channels (in the corresponding frequency band) not in Imprint (Fig.~\ref{fig:baseline} D `channel-specific’ baseline case). Ictal deviations were summarised per group by taking the median across channels. Finally, we computed the difference in MAD scores between the channel groups (MAD(\texttt{[in Imprint]}) – MAD(\texttt{[in LEM but not Imprint]})) to obtain a direct quantification of the Decision Point (channel-specific \textit{vs.} global baseline) effect.

We used a one-tailed paired-sample Wilcoxon rank-sum test to determine if ictal deviations were substantially higher for Imprint onset channels ($H_0:$ \texttt{[in Imprint]} = \texttt{[in LEM but not in Imprint]}, $H_1:$ \texttt{[in Imprint]} $>$ \texttt{[in LEM but not in Imprint]}). 

We repeated this post-hoc analysis for LEM. For each seizure we obtained a fixed, global, preictal baseline (115 to 10 seconds before marked onset) for each of the frequency bands from which we created a created global baseline distributions by pooling values from all channels (see Fig.~\ref{fig:baseline}B bottom). We captured ictal deviations by comparing early ictal activity (defined as the first five seconds of activity between clinically labelled onset time and LEM detection time) against this baseline using MAD scores. We divided onset channels into two groups: LEM onset channels (in the corresponding frequency band) \textit{vs.} Imprint onset channels not in LEM (Fig.~\ref{fig:baseline}D `global’ baseline case). Ictal deviations were summarised per group by taking the median across channels. Finally, we computed the difference in MAD scores between the channel groups (MAD(\texttt{[in LEM]}) – MAD(\texttt{[in Imprint but not LEM]})) to obtain a direct quantification of the Decision Point (channel-specific \textit{vs.} global baseline) effect. See Suppl. \ref{suppl:cg_methods} for further details on how ictal activity was scored against the baseline.

We used a one-tailed paired-sample Wilcoxon rank-sum test to determine if ictal deviations were substantially higher for LEM onset channels ($H_0:$ \texttt{[in LEM]} = \texttt{[in Imprint but not in LEM]}, $H_1:$ \texttt{[in LEM]} $>$ \texttt{[in Imprint but not in LEM]}).

\subsubsection*{Decision Point 2: Inclusion vs. exclusion of increases in low-frequency activity}
Next, we investigated whether inclusion or exclusion of increases in low-frequency activity impacted resultant onset channels. We compared EI and LEM, two algorithms that differ in their inclusion (LEM) or exclusion (EI) of increases in low-frequency activity. To enable an easier comparison, we focussed on onset channels identified by LEM and/or EI.

\textbf{Post-hoc analysis:} We identified nine independent sets of LEM onset channels, one for each of the nine frequency bands. We grouped these LEM onset channels using the definition of low- and high-frequency activity used in EI (i.e., lower or higher than 12.5~Hz). LEM onset channels exclusively detected in low-frequencies were labelled as $LEM_{LF}$. All other LEM onset channels (i.e., those exclusively detected in high-frequencies and those detected in both low- and high-frequencies were labelled as $LEM_{HF}$ (see Fig.~\ref{fig:low_freq}A\&B). 

We captured the proportion EI onset channels overlapping with $LEM_{LF}$ and $LEM_{HF}$ channels for each seizure (see Fig.~\ref{fig:low_freq} C\&D). The difference in overlap with $LEM_{HF}$ than \textit{vs.} $LEM_{LF}$ would form a direct quantification of the Decision Point effect (including \textit{vs.} excluding low frequency activity).

We first used a Fisher’s exact test to determine if overlap with EI is present for $LEM_{LF}$ and $LEM_{HF}$ channels.
For a subset of seizures with both $LEM_{LF}$ and $LEM_{HF}$ channels, we next used a one-tailed paired-sample Wilcoxon rank-sum test to determine if the overlap with EI was substantially higher for $LEM_{HF}$ channels ($H_0:$ Overlap between $LEM_{HF}$ and EI = Overlap between $LEM_{LF}$ and EI, $H_1:$ Overlap between $LEM_{HF}$ and EI $>$ Overlap between $LEM_{LF}$ and EI).

\subsubsection*{Decision Point 3: Inclusion vs. exclusion of electrodecrements }
Finally, we investigated whether inclusion or exclusion of electrodecrements impacted resultant onset channels. We compared Imprint and LEM, two algorithms that differ in their inclusion (LEM) or exclusion (EI) of electrodecrements. We algorithmically detected seizures where Imprint was capturing electrodecrements (see Fig.~\ref{fig:decrements}A). In these seizures, we focus on channels identified by either Imprint and/or LEM (see Fig.~\ref{fig:decrements}B). Further details on identification of electrodecrements is provided in Supplementary \ref{suppl:electrodec_methods}.

\textbf{Post-hoc analysis:} We selected seizures where at least one Imprint onset channel captured electrodecrements. We calculated the proportion of Imprint onset channels that were capturing electrodecrements. From this we assessed its correlation with the overlap between Imprint and LEM onset channels (see Fig.~\ref{fig:decrements}B). The correlation provides a direct quantification of the Decision Point effect (considering \textit{vs.} ignoring electrodecrements).

We used a Spearman’s rank correlation to determine if the overlap between onset channels was associated with the proportion of Imprint onset channels with electrodecrement ($H_0/H_1:$ there is not/is an association between \texttt{[proportion of Imprint onset channels with electrodecrement]} and \texttt{[overlap between Imprint and LEM onset channels]}). 

\subsection*{Code availability}
Analysis code will be made available at \url{https://github.com/SGascoigne97/} upon acceptance. 

\section*{Results}
Throughout these results, we use 100 seizures from 16 subjects as a bench-marking dataset. The three onset localisation algorithms, Imprint, LEM, and EI, were applied to all seizures.

\subsection*{Algorithms disagree on the presence and size of onset locations}
We first investigated whether each algorithm was able to identify onset channels. Of the 100 seizures, Imprint, LEM, and EI were unable to identify onset channels in one, 26, and 30 seizures, respectively. Comparing across algorithms, the seizure with no Imprint onset channels was also missing EI onset channels. A total of 12 seizures were missing both LEM and EI onset channels. Every seizure had onset locations detected by at least one algorithm (see Suppl. \ref{suppl:no_ons}). Seizures where LEM and/or EI were unable to locate onsets were spread across subjects - for only one subject, EI was unable to locate onsets across all of their seizures. 

Of the onset channels identified by each algorithm, LEM onset locations tended to be larger, with a median of 16.1\% of channels included (across all seizures) compared to Imprint and EI (median = 3.4\% and 5.1\% of channels, respectively). 

Finally, we investigated whether onset locations were consistent across algorithms. A minority of seizures ($1.8-2.9\%$) had the same or similar ($\geq75\%$ overlap) onset channels between two algorithms. No or minimal ($\leq25\%$) overlap occurred in $26.8-60\%$ of seizures. Where there was some overlap, Imprint and EI onset channels both tended to be a subset of the corresponding LEM onset channels, but LEM onset channels were rarely a subset of any other algorithm's onset channels. This result is consistent with the finding that LEM onset locations tended to be larger than those found by Imprint and EI.

\subsection*{Decision Point 1: Definition of baseline activity.}
To detect epileptic activity as something abnormal, we often first need to define what is normal; and in most instances this is some segment of baseline (preictal) data. We investigated the impact of different temporal (Decision Point 1a: fixed vs. moving baseline) and spatial (Decision Point 1b: global vs. channel-specific baseline) definitions for baseline activity. 

\subsubsection*{Decision Point 1a: Fixed vs. moving baselines impact onset locations}
\begin{figure}
    \centering
    \includegraphics[width=0.75\linewidth]{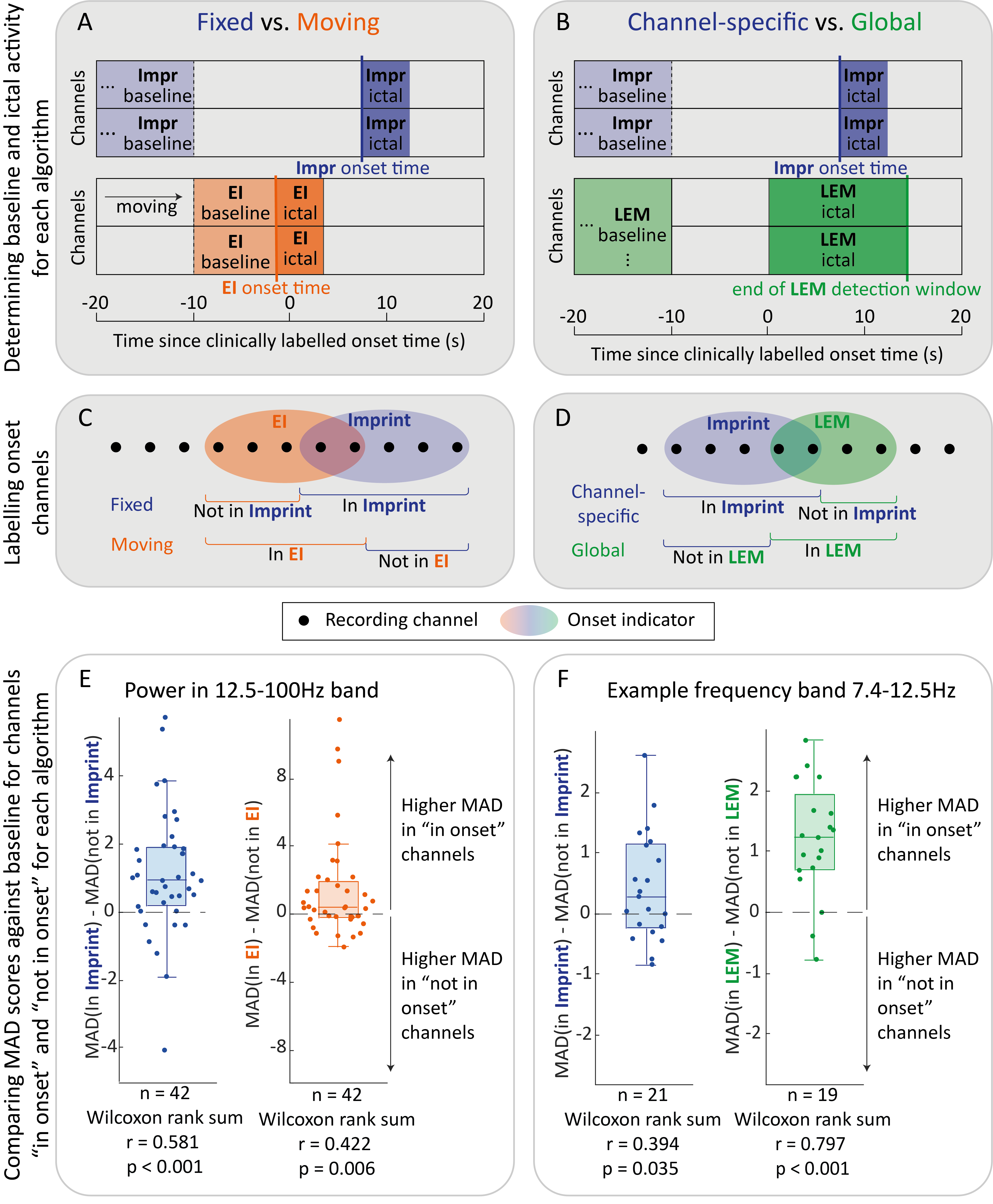}
    \caption[Selection of baseline (Decision Point 1a: fixed vs. moving and Decision Point 1b: channel-specific vs. global) substantially impacts onset locations.]{\textbf{Selection of baseline (Decision Point 1a: fixed vs. moving and Decision Point 1b: channel-specific vs. global) substantially impacts onset locations.} A\&B) Schematic showing example temporal location of baseline and ictal activity for each comparison. EI is shown in orange. 
    Ictal activity in each channel was then MAD scored against the respective baseline for each algorithm. 
    C\&D) Schematic showing onset locations for each pair of algorithms being compared and the subsequent classifications for each comparison. E\&F) Box plots showing the difference in MAD scores between Imprint onset channel and EI/LEM but not in Imprint (and vice-versa) based on classifications shown in panels C\&D. In the fixed vs. moving baseline comparison (E), we used broad-band high-frequency activity ($12.5-100$~Hz) as a consistent feature representing the activity captured by either algorithm. In the channel-specific vs. global comparison (F), we used band powers in seven frequency bands as consistent features and 7.4-12.5~Hz is shown here as an example. \textbf{Shorthand:} Imprint is shortened to Impr.}
    \label{fig:baseline}
\end{figure}

First, we investigated whether defining a fixed or moving baseline impacted resultant onset channels. To explore the implications of the time window used to calculate a baseline, we looked at Imprint (fixed baseline) and EI (moving baseline) onset locations (Fig.~\ref{fig:baseline}A).

To enable a direct comparison between the two algorithms, we used the same ictal activity filtered between 12.5-100~Hz band and scored it either against a fixed or moving baseline. We also focus only on Imprint and/or EI onset channels, and analyse why some were detected by one algorithm, but not the other. 

To ensure we could directly compare the same seizures between the two algorithms, we selected a subset of suitable seizures that showed different onset locations between Imprint and EI. A total of 42 seizures across 11 subjects were found to be suitable for this analysis. We excluded seizures with electrodecrements, EI onset times too early to create suitable baseline, or complete overlap between Imprint and EI onset channels.

First, we compared all Imprint onset channels \textit{vs.} EI onset channels not in Imprint (Fig.~\ref{fig:baseline}C `Fixed' baseline case). We found that early ictal activity showed larger deviation (as measured by MAD) from a fixed baseline in the Imprint onset channels (Fig.~\ref{fig:baseline}E left panel). Across the 42 seizures, we detected a large effect (Wilcoxon rank-sum $r = 0.581$, $p <0.001$). This indicates that Imprint is specifically detecting abnormalities, or deviations from a fixed baseline; and that EI onset channels not in Imprint simply did not contain sufficient deviations from a fixed baseline.

Second, we compared all EI onset channels \textit{vs.} Imprint onset channels not in EI (Fig.~\ref{fig:baseline}C `Moving' baseline case). We found that early ictal activity showed larger deviation (as measured by MAD) from a moving baseline in the EI onset channels (Fig.~\ref{fig:baseline}E right panel). Across the 42 seizures, we detected a moderate effect (Wilcoxon rank-sum $r = 0.422$, $p = 0.006$). This indicates that EI is specifically detecting abnormalities, or deviations from a moving baseline; and that Imprint onset channels not in EI simply did not contain sufficient deviations from a moving baseline.

Taken together, our results clearly show that fixed \textit{vs.} moving baselines have a large effect and indeed leads to disparate localisations.

\subsubsection*{Decision Point 1b: Channel-specific vs. global baselines impact onset locations}
Next, we investigated whether defining a channel-specific or global baseline impacted resultant onset channels. To explore the implications of the spatial information used to calculate a baseline, we looked at Imprint (channel-specific baseline) and LEM (global baseline) onset locations (Fig.~\ref{fig:baseline}B).

To enable a direct comparison between the two algorithms, we compared the same ictal activity across the same frequency bands. We also focus only on Imprint and/or LEM onset channels, and analyse why some were detected by one algorithm, but not the other. 

To ensure we could directly compare the same seizures between the two algorithms, we selected a subset of suitable seizures that showed different onset channels between Imprint and LEM. A total of 73 seizures across all 16 subjects were found to be suitable for this analysis. Analyses were performed independently for each of the frequency bands. 

The number of seizures remaining in the comparison for each frequency band ranged from four to 28 following exclusion of seizures with no LEM onset location in the given frequency band, seizures with electrodecrements, or complete overlap between Imprint and LEM onset locations. 

Of all nine frequency bands we investigated, we could only include seven in our analysis: no seizures had onset channels identified in the highest frequency band (101.9-132.5~Hz), and frequency band 4.4-7.4~Hz also could not be assessed as only four seizures had LEM onset locations in this frequency. 

First, we compared all Imprint onset channels \textit{vs.} LEM onset channels not in Imprint (Fig.~\ref{fig:baseline}D `channel-specific’ baseline case). We found that early ictal activity showed larger deviation (as measured by MAD) from a channel-specific baseline in the Imprint onset channels. In each of the frequency bands, we detected a moderate to large effect (Wilcoxon rank-sum $r >0.35$, $p <0.05$ for all seven frequency bands assessed). Fig.~\ref{fig:baseline}F visualises data for one example frequency band. This indicates that Imprint is specifically detecting abnormalities, or deviations from a channel-specific baseline; and that LEM onset channels not in Imprint simply did not contain sufficient deviations from a channel-specific baseline.

Next, we compared all LEM onset channels \textit{vs.} Imprint onset channels not in LEM (Fig.~\ref{fig:baseline}E `global’ baseline case). We found that early ictal activity showed larger deviation (as measured by MAD) from a global baseline in the LEM onset channels. In six of the seven frequency bands assessed, we detected a large effect (Wilcoxon rank-sum $r >0.6$, $p<0.005$). This indicates that LEM is specifically detecting abnormalities, or deviations from a global baseline; and that Imprint onset channels not in LEM simply did not contain sufficient deviations from a global baseline. See Supplementary \ref{suppl:gc_results} for results across other frequency bands. 

Taken together, our results clearly show that channel-specific \textit{vs.} global baselines have a large effect and indeed leads to disparate localisations.

See Supplementary \ref{suppl:gc_results} for additional effects of the amount of activity in the preictal segment on resultant onset channels. Across the seven frequency bands, preictal activity was higher in LEM onset channels not in Imprint than in Imprint onset channels for most seizures ($r > 0.6$, $p \leq 0.03$ for all frequency bands).

\subsection*{Decision Point 2: Inclusion vs. exclusion of increases in low-frequency activity impacts onset locations}
\begin{figure}
    \centering
    \includegraphics[width=0.75\linewidth]{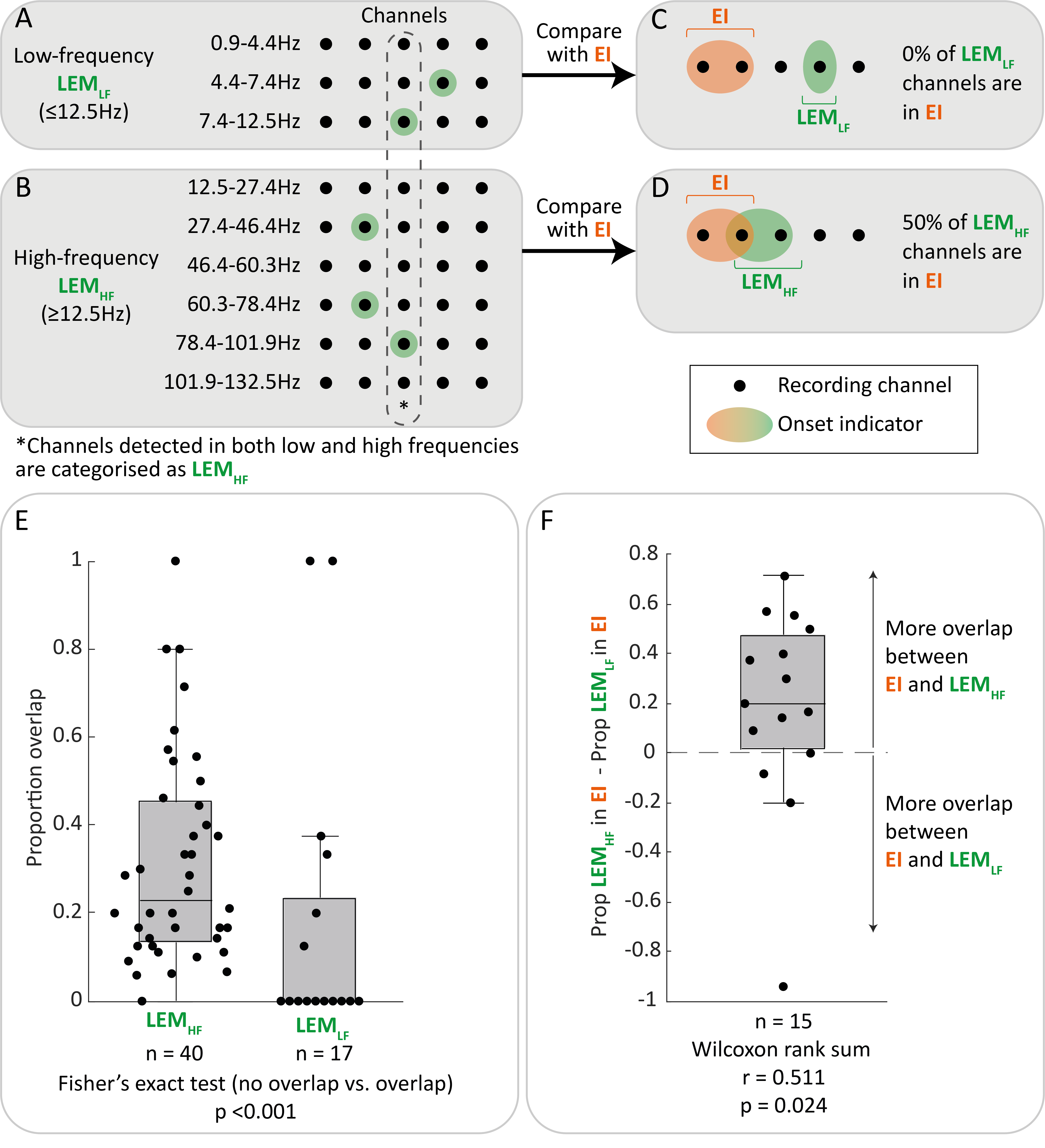}
    \caption[Inclusion or exclusion of increases in low-frequency activity (Decision Point 2) leads to substantially different 
 onset locations.]{\textbf{Inclusion or exclusion of increases in low-frequency activity (Decision Point 2) leads to substantially different 
 onset locations.} A\&B) Schematic showing how LEM onset channels are labelled as $LEM_{LF}$ or $LEM_{HF}$ based on the frequency band selected in LEM algorithm. LEM onset channels are circled in green for each frequency band. C\&D) Schematic of how LEM onset channels are compared with EI. LEM onset channels ($LEM_{LF}$ or $LEM_{HF}$) are circled in green; EI onset channels are circled in orange. E) Box plots showing comparisons of overlap with EI for $LEM_{LF}$ and $LEM_{HF}$ channels. Result of Fisher's exact test is presented. F) Box plot of difference in overlap between $LEM_{LF}$ and $LEM_{HF}$ channels ($LEM_{HF}-LEM_{LF}$). Only seizures with both $LEM_{HF}$ and $LEM_{LF}$ channels were included in this panel.}
    \label{fig:low_freq}
\end{figure}

Next, we investigated whether including or excluding increases in low-frequency activity impacted resultant onset channels. To this end, we compared LEM (includes increases in low-frequency activity) and EI (does not include increases in low-frequency activity) onset locations (Fig.~\ref{fig:low_freq}A-D).

We categorised LEM onset channels as low- ($LEM_{LF}$) or high-frequency ($LEM_{HF}$) based on their frequency band (see Fig.~\ref{fig:low_freq}A\&B). Next, we computed the proportion of EI onset channels that were contained within each category of LEM onset channels (Fig.~\ref{fig:low_freq}C\&D). 

To ensure we could directly compare the same seizures between the two algorithms, we selected a subset of suitable seizures that differed in onset channels between LEM and EI. A total of 42 seizures across 13 subjects were found to be suitable for this analysis after exclusion of seizures with missing LEM and/or EI localisations. 

First, we assessed whether LEM onset channel categories were associated with some or no overlap with EI (Fig.~\ref{fig:low_freq}E). We found that $LEM_{HF}$ tended to have some overlap with EI whilst $LEM_{LF}$ channels tended to have no overlap with EI (Fisher’s exact test: $p < 0.001$). 

Second, across 15 seizures with both $LEM_{LF}$ and $LEM_{HF}$ channels, we compared the difference in overlap metrics between channel categories (Fig.~\ref{fig:low_freq}F). We detected a large effect (Wilcoxon rank-sum $r = 0.511$, $p = 0.024$). 

Taken together, our results clearly show that inclusion \textit{vs.} exclusion of increases in low frequency activity in an onset localisation algorithm has a large effect and indeed leads to disparate localisations.

\subsection*{Decision Point 3: Inclusion vs. exclusion of electrodecrements impacts onset locations}
\begin{figure}
    \centering
    \includegraphics[width=0.75\linewidth]{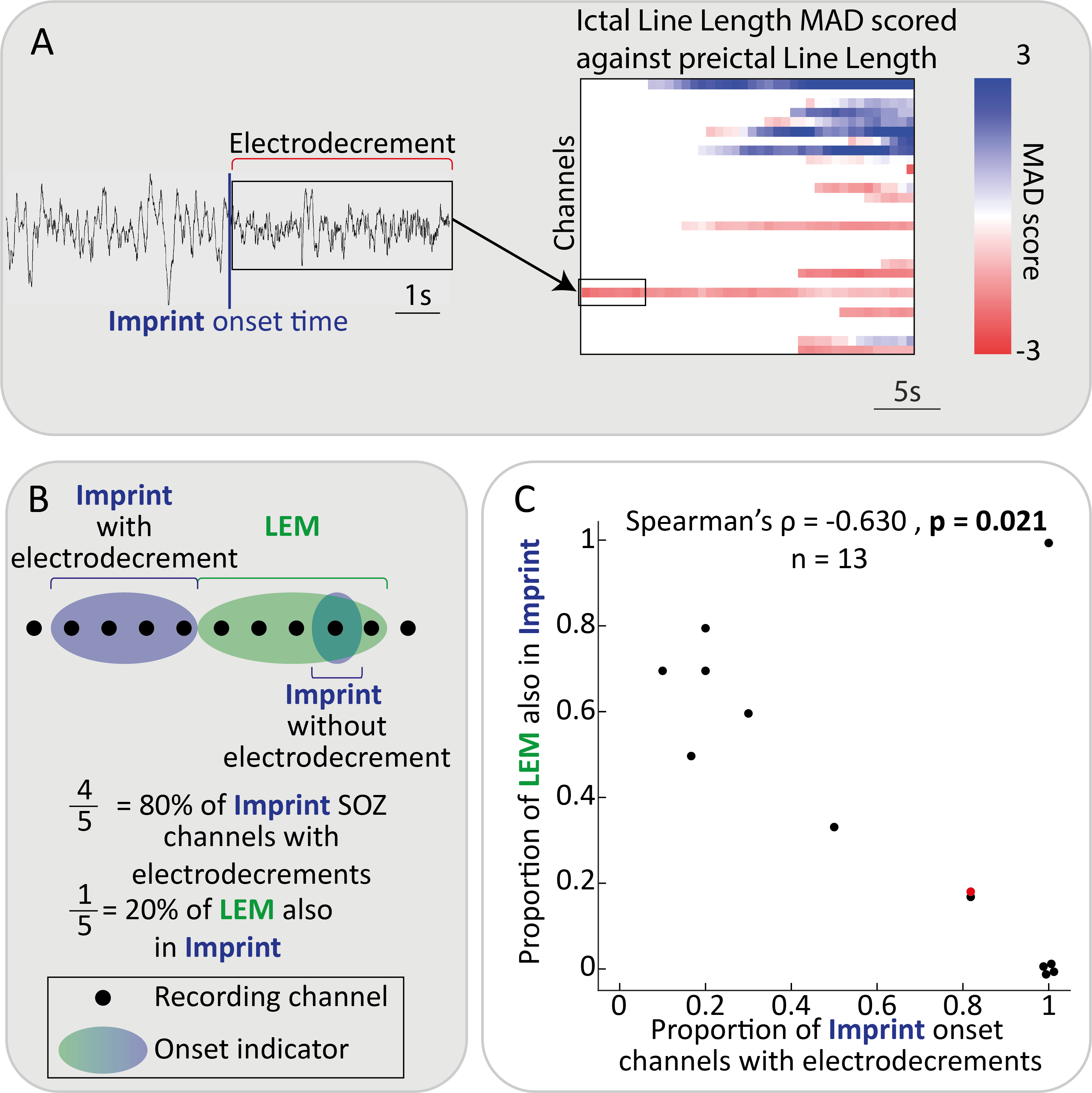}
    \caption[Inclusion or exclusion of electrodecrements (Decision Point 3) leads to substantially different onset locations.]{\textbf{Inclusion or exclusion of electrodecrements (Decision Point 3) leads to substantially different onset locations.} A) Time series for an example channel with electrodecrements detected by the Imprint algorithm (left). Heat-map of Median Absolute Deviation (MAD) scores between preictal baseline and early ictal activity for the example seizure (right). Red is used to indicate decrements in activity (electrodecrements), captured as negative MAD scores. Only time points where activity was highlighted by the Imprint algorithm are shown here. B) Schematic showing the comparison between LEM and Imprint, where Imprint onset channels were separated into those with and without electrodecrements. C) Scatterplot of the proportion of Imprint onset channels capturing electrodecrements and the overlap between Imprint and LEM onset locations. Example seizure shown in panel B is shown in red.}
    \label{fig:decrements}
\end{figure}

Finally, we investigated whether including or excluding electrodecrements impacted resultant onset channels. To this end, we compared Imprint (includes electrodecrements) and LEM (does not include electrodecrements) onset locations (Fig.~\ref{fig:decrements} A\&B).

For each seizure, we computed the proportion of Imprint onset channels capturing electrodecrements. To assess whether the overlap between Imprint and LEM onset locations is mediated by the proportion of Imprint onset channels capturing electrodecrements, we computed a Spearman’s rank correlation between the proportion of Imprint onset channels with electrodecrements and the proportion of LEM onset channels also in Imprint.

We selected seizures where we captured electrodecrements. A total of 13 seizures across seven subjects were found to be suitable for this analysis after exclusion of seizures with no LEM channels, or no Imprint channels after the removal of electrodecrements (see Suppl.~\ref{suppl:electrodec_methods}). 

First, in Imprint onset channels, the median proportion of channels with electrodecrements was 90.9\% (IQR 25- 100\%), suggesting that when Imprint onsets capture electrodecrements, such activity dominate the vast majority of onset channels.

We then assessed whether overlap between Imprint and LEM onset locations was associated with the proportion of the Imprint onset channels that included electrodecrements. We found a substantial negative relationship between the former and the latter (Fig.~\ref{fig:decrements}C, Spearman’s rank: $\rho = -0.630$, $p = 0.021$).

We further assessed the change in overlap between Imprint and LEM having removed electrodecrements from the Imprint. There was an increase in the proportion for most seizures (see Suppl.~\ref{suppl:electrodec_results}), further supporting our previous result. 

Taken together, our results clearly show that inclusion \textit{vs.} exclusion of electrodecrements in an onset localisation algorithm has a large effect and indeed leads to disparate localisations.

\subsection*{Further Considerations}
There were a number of additional considerations that did not form Decision Points -- these were explored as supplementary analyses. 

We qualitatively assessed the impact of thresholding in activity detection in a subject with visually similar seizure onsets. Using the Imprint and EI algorithms, we highlighted changes in activity on the icEEG time series and assessed if channels were excluded from onset locations due to thresholding. This demonstrated that threshold selection can affect within-subject consistency of onset locations. See Supplementary~\ref{suppl:thresh_abno_results} for results.

The LEM algorithm implements an entropy threshold to limit the size of the onset location. For the seizures with no LEM onset location, we relaxed the entropy threshold until we found an onset location. We found that increasing the entropy threshold resulted in larger onset locations (i.e., more onset channels; $p<0.001$). See Supplementary~\ref{suppl:entropy_thresh} for results.  

Finally, we assessed the within-subject concordance of onset locations using each algorithm. Within-subject concordance tended to be low ($d<0.5$) for all algorithms. However, within-subject concordance across seizures is not necessarily expected to be high as time varying modulators can impact the type an location of activity at seizure onset. See Supplementary~\ref{suppl:concordance} for results.

\section*{Discussion}
This study explored the differences in onset locations across three automatic onset localisation algorithms using intracranial EEG data. We found low overlap between the onset locations identified by the three algorithms, suggesting that no two algorithms capture activity in the same way. Differences between onset locations across algorithms were mapped back to decisions made in the development of the algorithms. Each Decision Point had a notable effect on the subsequent onset locations; however, no singular Decision Point was responsible for all differences in onset locations detected across algorithms. 

The results presented here can be interpreted as a word of caution against using a blanket approach to algorithmically detect onset locations, without taking into account the underpinnings of the algorithms used. Instead, it is crucial to consider the type of activity that is of interest, bearing in mind that activity at seizure onset can vary across and within individuals \cite{pelliccia2013ictal, salami2020seizure, Donos2018, velascol2000functional, perucca2014intracranial}. This information can then aid in the selection of an algorithm best placed to detect the activity type - taking our Decision Points as a guide.  

Onset localisation algorithms are often created and assessed within the context of surgical outcome prediction, where the onset location is compared against the surgical resection \cite{megevand2014electric, Weiss2015SeizureStudy, li2021neural,klimes2024interictal,Diamond2023InterictalSource}, and/or clinically labelled onset locations \cite{murin2018sozrank,pellegrino2016source,van2013ictal}.

In this work, we did not compare onset locations against clinically labelled onset locations (e.g., \cite{andrzejak2015localization}) or resections as we aimed to assess quantitative differences between onset localisation algorithms rather than to determine the `best' method. Instead, we highlighted how the algorithms differ and the resultant impact on the channels selected by each algorithm. This work could be extended by assessing the impact of each Decision Point with respect to overlap with clinically labelled onset locations and/or resections. 

Our work had several limitations. The dataset is relatively small and single centre. Future work could apply these methods using a larger cohort from multiple centres. This data does not include metadata which should be used to determine if subject demographics (e.g., epilepsy type, age, etc.) impact our findings. However, to yield reliable results, this would require a larger dataset with good representation. 

The algorithms compared within this work were designed specifically to localise the channels involved in seizure onset rather than to detect seizures in continuous data (e.g., \cite{kharbouch2011algorithm, ahammad2014detection}). Future work could apply similar methods to seizure detection algorithms to determine which Decision Points impact the time and localisation of the onsets detected. 
Additionally, we did not compare all algorithms in the literature, but compare three leading examples, representative of the wider field. Future work could also apply this approach to other onset localisation algorithms (e.g., \cite{Weiss2015SeizureStudy, van2011accurate, van2013ictal}). 

Automatic onset localisation algorithms are not commonly used in clinical settings \cite{ahufinger2019user}. Creating a consumer-friendly user interface (UI) will improve accessibility and uptake \cite{yuan2013evaluation}. Such a UI could display icEEG signals with the onset location labelled, allowing for quick validation of the onset location. Further, future work could use an interactive tool to visualise changes in the onset detected alongside decisions made whilst designing an algorithm. 

In summary, each algorithm offers unique insights, however there is no one-size-fits-all approach. Instead, an approach combining existing algorithms, along with a deeper understanding of their underlying methodologies, could improve confidence and reliability of onset locations. Addressing these challenges will require the collaboration of clinicians, researchers, and engineers to refine these tools and enhance their clinical utility.

\end{doublespace}

\section*{Data Availability}
Data was obtained from short-term SWEZ-ET~Hz publicly available database \cite{burrello2019laelaps} (accessible at \url{http://ieeg-swez.et~Hz.ch/}). 

\section*{Author Contributions}
\begin{itemize}
  \item Conceptualization: SJG MVV NE PNT ATC YW
  \item Methodology: SJG MVV NE YW
  \item Software/validation: SJG MVV NE BS ATC YW
  \item Formal analysis: SJG YW
  \item Writing: SJG MVV NE CT HW BS KW PNT YW
  \item Supervision: ABDS RHT KW PNT ATC YW
\end{itemize}

\section*{Acknowledgements and Funding}
We thank members of the Computational Neurology, Neuroscience \& Psychiatry Lab (www.cnnp-lab.com) for discussions on the analysis and manuscript. S.J.G and H.W are supported by the Engineering and Physical Sciences Research Council (EP/L015358/1) and ADLINK; N.E is supported by Epilepsy Research Institute UK (OSR/0550/ERUK); P.N.T and Y.W are both supported by UKRI Future Leaders Fellowships (MR/T04294X/1, MR/V026569/1). A.T.C and M.V.V are supported by grant PID2020-119072RA-I00, funded by MCIN/AEI/10.13039/501100011033. M.V.V is supported by grant PTQ2022-012679, funded by MCIN/AEI/10.13039/501100011033.

\section*{Competing interests}
None of the authors have any conflicts of interest to disclose. 

\newpage

\bibliography{refs}

\newpage


\renewcommand{\thefigure}{S\arabic{figure}}
\renewcommand{\thetable}{S\arabic{table}} 
\counterwithin{figure}{subsection}
\counterwithin{table}{subsection}
\renewcommand\thesection{S\arabic{section}}
\setcounter{section}{0}
\setcounter{subsection}{0}

\section*{Supplementary}
\subsection*{Glossary of acronyms}
\subsubsection*{Clinical terms}
\begin{itemize}[label={}]
    \item \textbf{EEG:} Electroencephalography
    \item \textbf{icEEG:} Intracranial EEG
\end{itemize}

\subsubsection*{Statistical terms}
\begin{itemize}[label={}]
    \item \textbf{MAD:} Median Absolute Deviation
\end{itemize}

\section{Supplementary Methods \label{suppl:suppl_meth}}
\subsection{Harmonising algorithms} \label{suppl:harmon}
In this work, we aim to highlight the differences between onset localisation algorithms as a result of methodological differences. It is not possible to exhaustively compare all differences between algorithms, therefore we harmonised the algorithms such that our results best highlight the impact of each Decision Point with minimal additional factors to consider. These harmonisations did not change any underpinnings of the algorithms.

The LEM algorithm uses a 30 second window to identify activity from which the onset location is found, from the LEM detection time, the algorithm then looks at earlier activity to find the onset location. As Imprint requires that activity persists across a nine-second window, we implemented an exclusion criterion that any Imprint onset location with an onset time later than 20 seconds after the clinically labelled onset time would be removed from further analysis. This same criteria was applied to EI onset locations. Such a late onset time is likely a result of algorithms detecting propagated activity rather than the onset of the seizure. 

The Imprint algorithm uses a 10 second ictal buffer to account for potentially mislabelled onset times. Specifically, the preictal baseline excludes the last 10 seconds and the time window in which ictal activity can be found begins 10 seconds before the clinically labelled onset time. Therefore, we introduced the same ictal buffer to the implementation of the EI algorithm. The LEM algorithm excluded the last 10 seconds of the preictal segment from the baseline but did not adjust the point of earliest detection. Figure ~\ref{fig:baseline} A\&B demonstrate the use of the ictal buffer across algorithms. 

\subsection{Thresholding Abnormalities}\label{suppl:thresholds_methods}
All algorithms have parameters and thresholds that must be determined by the user. For example, EI computes an intermediary value UN which estimates the change in activity up to the current time step.

The Imprint algorithm uses a threshold to identify time points where there is potential seizure activity. The LEM algorithm uses two thresholds, applied on the variables GA and AE, to identify time-frequency windows with power increases confined to a reduced number of electrode contacts. Imprint and LEM thresholds (MAD, and GA and AE, respectively) were set by assessing robustness of onset locations across thresholds using an alternative icEEG data set. 

We selected thresholds for each algorithm by scanning potential threshold values and assessing the changes in onset location size and onset time. This was done using datasets independent of the data used in this study. We assessed the robustness of onset locations across thresholds to determine if the onset locations detected were sensitive to slight changes in thresholds. For the MAD threshold used in Imprint, we scanned values from one to five, in steps of 0.5, and recomputed onset locations for all seizures. We captured the number of onset channels, onset time, and computed Cohen’s kappa between onset locations with one step in threshold. Additionally, we validated the UN threshold of 15 as previously suggested (Diamond et al., 2023).

Parameters and thresholds are set in each algorithm:
\begin{itemize}
    \item EI: EI threshold (EI$\geq$0.3 is included in the onset location) 
    \item Imprint: MAD $\geq$3 is labelled as abnormal
    \item LEM: Time-frequency windows with global activation (GA) percentile $\geq$83 and activation entropy (AE) $\leq$0.5 are included in the onset location.
\end{itemize}

\subsection{Exploring differences in onset locations across algorithms} \label{suppl:cat_seizures}
\begin{figure}
    \centering
    \includegraphics[width=0.75\linewidth]{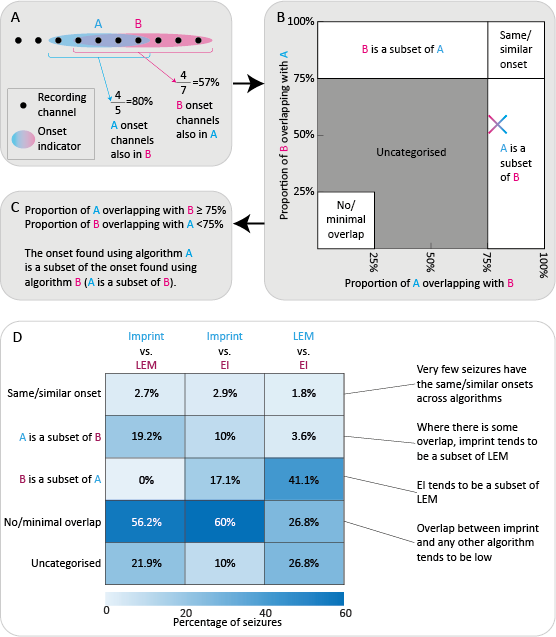}
    \caption[Workflow of comparisons of onset locations across algorithms.]{\textbf{Workflow of comparisons of onset locations across algorithms.}A) Schematic showing comparison of example onset locations A and B (left) through isolation of channels in each onset location (middle) and computation of the overlap metric i.e., proportion of channels with overlapping with the other algorithm's onset location (right). B) Schematic showing how seizures were categorised for each pair of algorithms using 25\% and 75\% thresholds. C) Label for seizure given overlap metrics. D) Heatmap of the percentage of seizures in each category for each pair of algorithms.}
    \label{fig:suppl_comparisons}
\end{figure}

First, we investigated whether onset locations were consistent across algorithms. Per pair of algorithms, we computed the proportion of each onset location captured within the other in both directions (see Fig.~\ref{fig:suppl_comparisons}A). As the size of the onset locations detected could vary across algorithms, we computed this metric in both directions such that we can capture instances where one algorithm's onset location is a subset of another's (see Figure \ref{fig:suppl_comparisons}B\&C). 

\subsection{Decision Point 1: Definition of baseline activity} \label{suppl:baseline_methods}
In both comparisons we summarised the change in activity from baseline using sliding one-second windows with 7/8th second overlap, for each channel. We used median absolute deviation (MAD) scores to compute the change in activity from baseline to early ictal for the algorithms included in each comparison. MAD scores for each channel were summarised across the selected ictal period using the median, we then summarised across channel groups for each comparison.

Channels were classified based on their presence in one or both onset locations, as these analyses compared the change in activity from baseline assuming that the channel had seizure activity at onset, channels that were not present in either algorithm's onset location were excluded.

\subsubsection{Decision Point 1a: Fixed vs. moving baselines} \label{suppl:fm_methods}
Here we compared Imprint and EI onset locations as they both use channel-specific baselines, but differ in the time windows used to create them. We compared the change in activity from baseline to the first changes seen in the ictal period.

For each seizure, we divided onset channels into two groups: Imprint onset channels \textit{vs.} EI onset channels not in Imprint 

We investigated the differences in onset times between algorithms to determine if we could use the same activity as our `early ictal’ period. A difference in onset times between algorithms is sufficient evidence to suggest that each comparison should use the ictal period defined by the algorithm being tested. 

As the Imprint and EI algorithms use different features to detect activity, we created a pseudo-marker that could capture activity that would be captured by both algorithms. Under the assumption that the EI algorithm is more strongly influenced by increases in high-frequency activity rather than decreases in low-frequency activity, we chose band power in broad-band high-frequency activity (13-100~Hz) as our marker for comparing changes in activity in Imprint and/or EI onset channels.

For each seizure we computed a fixed (120 to 10 seconds before clinically labelled onset time and moving (up to 10 seconds before EI onset time, excluding activity that is in the preictal segment) baseline for each onset channel (see Fig.~\ref{fig:baseline}). We MAD scored early ictal activity (first five seconds of activity following the Imprint or EI onset time) against the corresponding baseline; summarising across time using the median. 

We excluded seizure with an EI onset time earlier than 5 seconds into the detection window as creating a baseline with less than 5 seconds of data was insufficient to create a robust distribution.

We expect to see a larger change from preictal activity in Imprint onset channels compared to EI (not Imprint) onset channels, equally we expect that change from the moving baseline will be larger in EI onset channels than in Imprint (not EI) onset channels. 

\subsubsection{Decision Point 1b: Channel-specific vs. global baselines} \label{suppl:cg_methods}

Here we compared the Imprint and LEM algorithms as both use a consistent baseline but differ in whether this baseline was created across channels (LEM) or for each channel individually (Imprint). 

As the LEM algorithm looks at each feature individually and highlights features within which potential seizure activity is detected, we repeated our analysis for each of the nine frequency ranges used by LEM. A seizure was excluded if there were no onset channels associated with the selected frequency. 

We computed a channel-specific and a global baseline (115 to 10 seconds before clinically labelled onset time) and early ictal activity as defined by the LEM algorithm (activity between clinically labelled onset time and the LEM detection time). We computed the median MAD across LEM onset channels (including overlap) or Imprint only. 

We hypothesised that change from a channel-specific baseline will be larger in Imprint onset channels than in LEM only channels and that change from a global preictal will be larger in LEM onset channels than in Imprint only channels. We additionally expect that channels with a high preictal activity will be included in LEM but not Imprint. 

\subsubsection{Decision Point 3: Inclusion vs. exclusion of electrodecrements} \label{suppl:electrodec_methods}
Per onset channel, we identified time windows with electrodecrements in activity as those with MAD scores below -1 in four or more features and no MAD scores greater than three in any of the remaining features. If at least 10\% of time windows were found to be decreasing, this channel was labelled as ‘captures electrodecrements’. 

\section{Supplementary Results}
\subsection{Algorithms disagree on the presence and size of onsets} \label{suppl:no_ons}
We investigated the spread of seizures with missing onsets across subjects for Imprint, LEM and EI. There was only one seizure with a missing Imprint onset location. Missing LEM onsets were present in seven of the 16 subjects included in this study. No subject had all onsets missing using LEM. The proportion of seizures with missing LEM onsets varied from 7\% to 69\%. Missing EI onset location were present in ten of the 16 subjects included in this study. One subject (ID15) had no EI onset locations found. The proportion of seizures with missing EI onsets varied from 7\% to 100\%.

When considering missing onsets between algorithms, there were no seizures with missing Imprint and LEM onsets \ref{tab:missed_L_I}, 12 seizures were missing both EI and LEM onsets \ref{tab:missed_L_E}, and one seizure was missing an EI and Imprint onset \ref{tab:missed_E_I}.

\begin{table}[h]
    \centering
    \begin{tabular}{|c|cc|}
    \hline
    & LEM & No LEM \\
    \hline
    Imprint & 73 & 26 \\
    No Imprint & 1 & 0 \\
    \hline
    \end{tabular}
    \caption{Table of the number of missed onset locations between LEM and Imprint algorithms.}
    \label{tab:missed_L_I}
\end{table}

\begin{table}[h]
    \centering
    \begin{tabular}{|c|cc|}
    \hline
    &LEM&No LEM\\
    \hline
    EI&56&14\\
    No EI&18&12\\
    \hline
    \end{tabular}
        \caption{Table of the number of missed onset locations between LEM and EI algorithms.}
    \label{tab:missed_L_E}
\end{table}

\begin{table}[h]
    \centering
    \begin{tabular}{|c|cc|}
    \hline
    &EI&No EI\\
    \hline
    Imprint&70&29\\
    No imprint&0&1\\
    \hline
    \end{tabular}
    \caption{Table of the number of missed onset locations between EI and Imprint algorithms.}
    \label{tab:missed_E_I}
\end{table}

\subsection{Decision Point 1: Definition of baseline activity}
\subsubsection{Decision Point 1b: Channel-specific vs. global baselines impact onset locations} \label{suppl:gc_results}
There were 73 seizures with both Imprint and LEM onset locations found. We created a table of seizures with an Imprint onset location (excluding seizures with decreases) and a LEM onset location for each of the nine frequency bands. Finally, we excluded any seizures with complete overlap between Imprint and LEM for each frequency band. The number of seizures remaining in the comparison ranged from four to 28. No seizures had onset channels identified in the highest frequency band (101.9 - 132.5~Hz). The second frequency band (4.4-7.4~Hz) also could not be assessed as only four seizures had LEM onset locations detected in this frequency.

Finally, we performed one-tailed paired-sample Wilcoxon rank sum tests to determine if there was a difference in the change in activity in different channel groups. When comparing against the channel-specific baseline, the change in activity was higher for Imprint onset channels than it was for LEM onset channels not in Imprint ($r>0.35$, $p<0.05$ for all seven frequency bands). When comparing against the global baseline, the change in activity was higher for LEM onset channels than it was for Imprint onset channels not in LEM ($r>0.6$, $p<0.005$ for all seven frequency bands).

We additionally compared the amount of activity in the preictal segment under the hypothesis that channels with more preictal activity would be more likely to be included in the LEM onset location. Across the seven frequencies assessed, preictal activity was higher in LEM onset channels not in Imprint compared to Imprint onset channels for most seizures ($r>0.6, p\leq0.03$ for all frequency bands).

\begin{figure}
    \centering
    \includegraphics[width=0.75\linewidth]{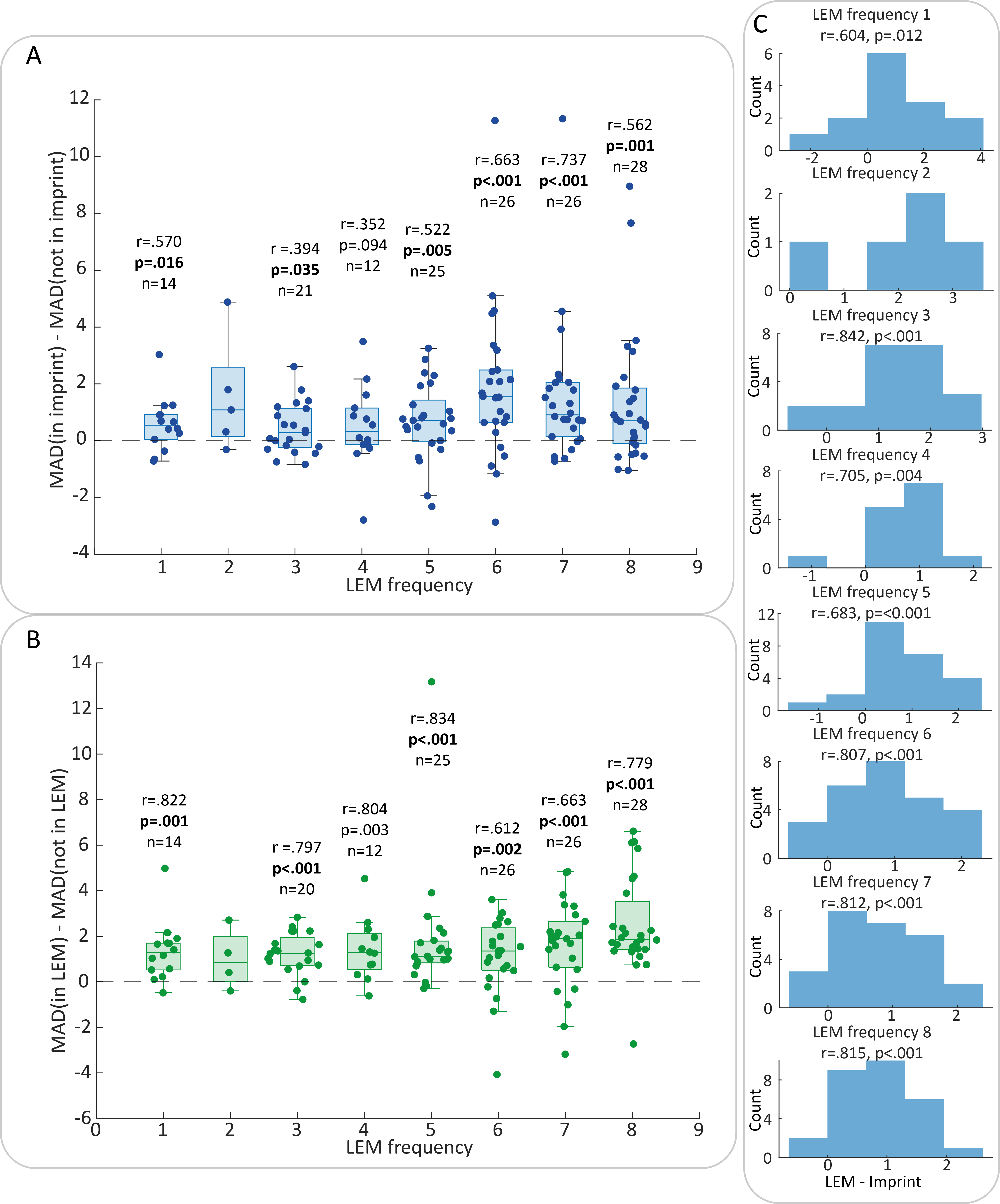}
    \caption[Selection of channel-specific or global baseline is substantially associated with differences in onset locations between Imprint and LEM for most frequency bands.]{\textbf{Selection of channel-specific or global baseline is substantially associated with differences in onset locations between Imprint and LEM for most frequency bands.}A) Box plots showing the difference in MAD scores between Imprint onset channels and LEM onset channels not in Imprint across eight of the frequency bands. B) Box plots showing the difference in MAD scores between LEM onset channels and Imprint onset channels not in LEM across eight of the frequency bands. In panels A and B, frequency nine has no associated box plot as this frequency had no onset channels across all seizures. No statistics are presented for frequency two as the sample size was too small to perform analysis. C) Histograms of difference in preictal activity between LEM onset channels not in Imprint and Imprint onset channels (LEM - Imprint).}
    \label{fig:enter-label}
\end{figure}
\subsection{Decision Point 3: Inclusion vs. exclusion of electrodecrements impacts onset locations} \label{suppl:electrodec_results}

For all seizures with at least one decreasing onset channel we recomputed onset locations following the identification and removal of all decreasing time windows (see Fig\ref{fig:decrements}A\&B). The updated onset locations were then compared with LEM onset locations. We expect that overlap between Imprint and LEM will improve having removed electrodecrements from Imprint onset locations.

\begin{figure}
    \centering
    \includegraphics[width=0.75\linewidth]{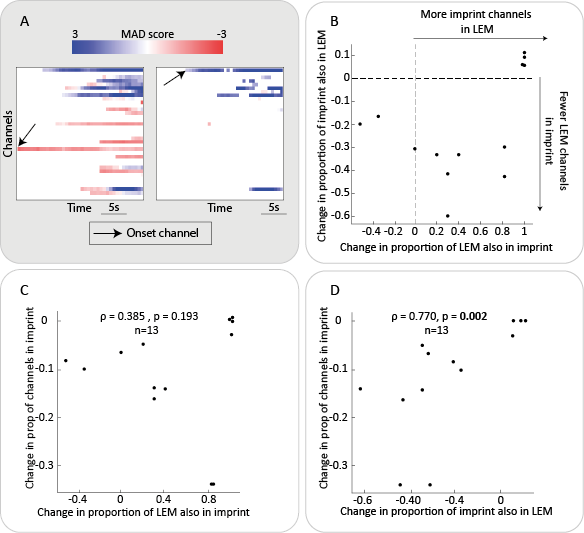}
    \caption[Removing electrodecrements from Imprint onsets is associated with an increase in the proportion of LEM onset channels that are also in Imprint.]{\textbf{Removing electrodecrements from Imprint onsets is associated with an increase in the proportion of LEM onset channels that are also in Imprint.}A) Heatmaps of median absolute deviation (MAD) scores of ictal line length against a channel-specific preictal baseline for a channel capturing electrodecrements (right), and the same heatmap following the exclusion of electrodecrements (left). Onset channels for each case are indicated with arrows. B) Scatterplot of the change in proportion of Imprint onset channels also in LEM when removing electrodecrements from the Imprint  (after-before). C) Correlation between the change in the proportion of LEM onset channels also in Imprint and the change in Imprint size when removing electrodecrements (after-before). D) Correlation between the change in the proportion of Imprint onset channels also in LEM and the change in Imprint size when removing electrodecrements (after-before).}
    \label{fig:suppl_dec}
\end{figure}

Overlap between LEM and Imprint increased after excluding electrodecrements in ten of the 13 seizures included. The proportion of Imprint in LEM remained constant in five seizures and decreased in three (see Fig.~\ref{fig:suppl_dec}B). Changes in proportion of LEM in Imprint were correlated with changes in Imprint onset location size after removal of electrodecrements ($\rho=0.77, p=0.002$), this relationship was not consistent when comparing against changes in the proportion on LEM in Imprint ($\rho=0.385, p=0.193)$. (see Suppl. Fig.~\ref{fig:suppl_dec}C\&D))

\subsection{Thresholding abnormalities} \label{suppl:thresh_abno_results}
\begin{figure}
    \centering
    \includegraphics[width=0.75\linewidth]{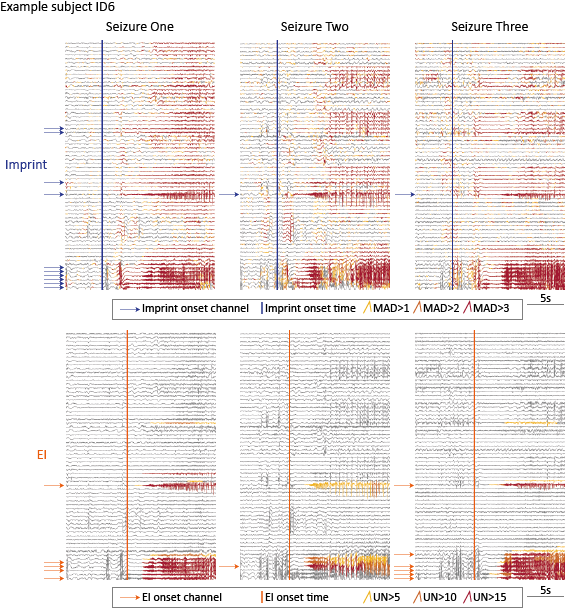}
    \caption[icEEG time series for three example seizures with Imprint and Epileptogenicity Index (EI) onset locations.]{\textbf{icEEG time series for three example seizures with Imprint and Epileptogenicity Index (EI) onset locations.} The Imprint plots (top) show time windows where MAD scores exceed one, two, and three as shown in yellow, orange, and red respectively. An MAD score of three was required for activity to be highlighted by the algorithm. The Imprint onset times are shown in blue. The EI onset times are shown in orange.
    Imprint onset channels are indicated with blue arrows.
    The EI plots (bottom) show the same time series with time windows highlighted based on the UN value. UN values are shown in yellow, orange, and red if the UN exceeded five, 10, and 15, respectively. The EI algorithm required activity to exceed a value of 15 to be detected by the algorithm. The EI onset times are shown in orange. EI onset channels are indicated with orange arrows.}
    \label{suppl_fig:thresholds}
\end{figure}

All algorithms require thresholds to distinguish between normal and pathological activity, here we performed a visual assessment of the thresholds used in the EI and Imprint algorithms and the resultant impact on onset locations. Figure \ref{suppl_fig:thresholds} displays Imprint and EI onset locations for one example subject with similar onsets based on visual assessment. EEG time series were coloured to reflect the value which was compared against a threshold for each algorithm: MAD score for Imprint and UN for EI. Time windows shown in red exceeded the thresholds used by the algorithms, and thus were sufficiently abnormal to be detected within the algorithm. 

\subsection{Exclusion of widespread activations in LEM} \label{suppl:entropy_thresh}
\begin{figure}
    \centering
    \includegraphics[width=0.75\linewidth]{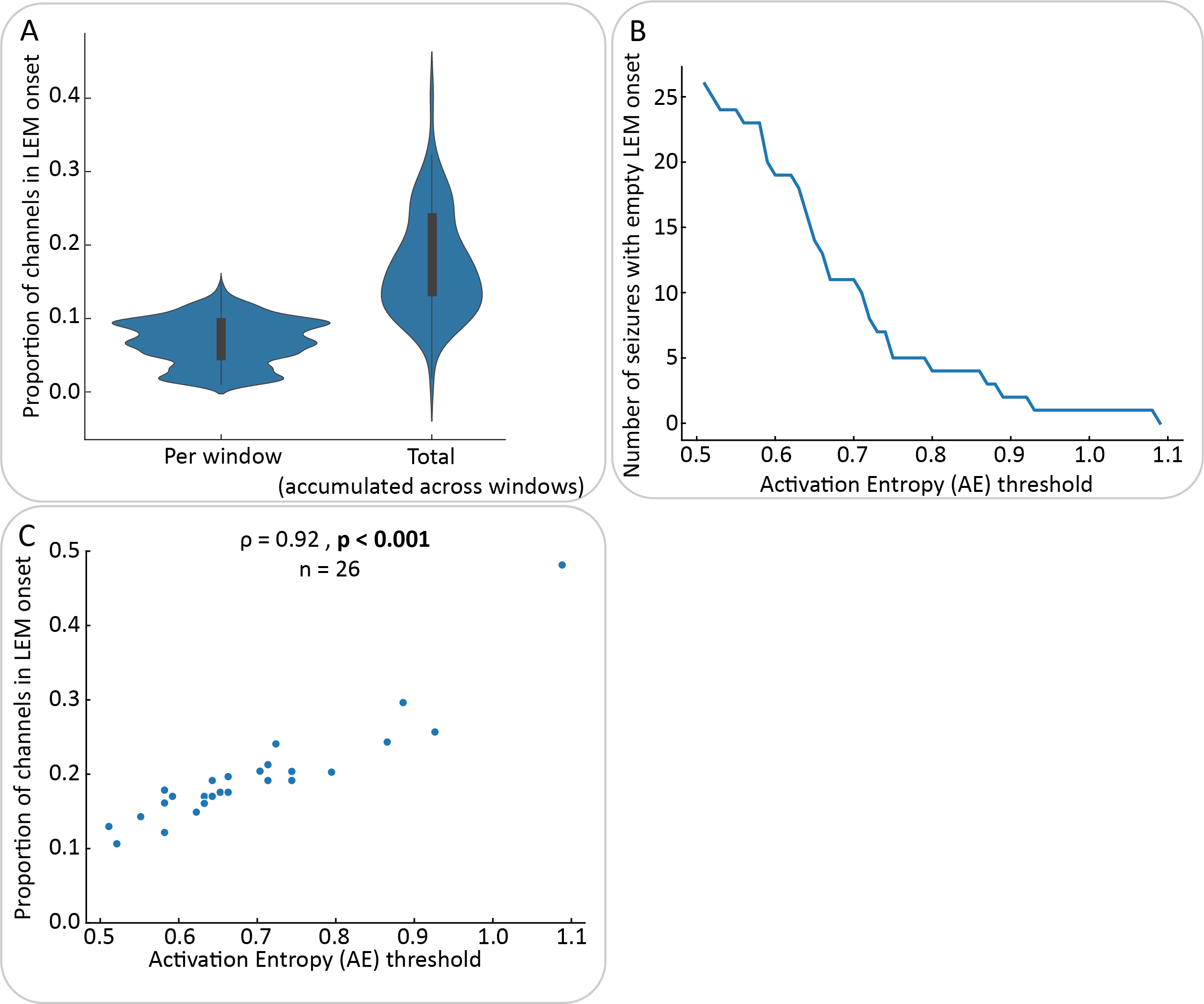}
    \caption[The LEM algorithm uses an entropy threshold to limit the spread of activations associated to each time-frequency window that can be included in the onset location.]{\textbf{The LEM algorithm uses an entropy threshold to limit the spread of activations associated to each time-frequency window that can be included in the onset location.}A) Violin plots showing the proportion of onset channels  across seizures at each detected time-frequency window (left) and accumulated across all detected time-frequency windows (right).
    For the AE threshold used in this study (0.5), 26 seizures were missing a LEM onset location. The remaining 74 seizures had a median of 16.1\% of onset channels with an inter-quartile range of 10\%.
    B) The number of seizures with missing onset location decreased as we gradually increased the entropy threshold.
    C) Correlation between the entropy threshold which first yielded an onset location and the proportion of channels in this onset location across the 26 seizures with initially missing LEM onset locations (Spearman’s rank: $\rho = 0.92, p < 0.001$).}
    \label{supplfig:aethreshold}
\end{figure}

The LEM algorithm uses an entropy threshold that limits the spatial spread of activations at each detected time-frequency window. In the 26 seizures with missing LEM onsets, we gradually relaxed the entropy threshold until an onset location was detected (see Fig.~\ref{supplfig:aethreshold}B). There was a substantial positive correlation between the entropy threshold which yielded an onset location proportion of channels in this onset location (Spearman’s rank: $\rho = 0.92, p < 0.001$) across the 26 seizures with initially missing LEM onset locations (see Fig.~\ref{supplfig:aethreshold}C).

Automatic methods can identify much larger onset locations than visual localisation, therefore the inclusion of a threshold to exclude seizures with onset locations that are too large can act as a measure preventing detection of propagated activity as onset if earliest change in activity was not captured by the algorithm, as is the case in the LEM algorithm. 

\subsection{Concordance within subjects} \label{suppl:concordance}
\begin{figure}
    \centering
    \includegraphics[width=0.5\linewidth]{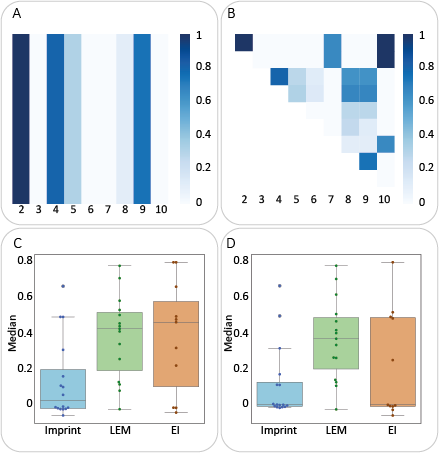}
    \caption[There is weak within-subject concordance between onsets using the same algorithm across seizures.]{\textbf{There is weak within-subject concordance between onsets using the same algorithm across seizures.}A) Ordered concordance for subject ID12 using the Imprint algorithm. B) Pairwise concordance for subject ID12 using the Imprint algorithm. C) Box-plots of median ordered concordance per subject for all three algorithms. D) Box-plots of median pairwise concordance per subject for all three algorithms.}
    \label{supplfig:concordance}
\end{figure}

We investigated the variability in seizure onset locations within individuals for each algorithm using Cohen's $\kappa$. We assessed ordered concordance (i.e., concordance between seizures in chronological order), and pairwise concordance (i.e., concordance between each pair of seizure irrespective of order). See Figure \ref{supplfig:concordance}A and B for ordered and pairwise concordance values for an example subject, respectively. We computed the median concordance for each subject in each algorithm to compare across algorithms (see Fig.~\ref{supplfig:concordance}C and D). Concordance between seizures was notably lower using the Imprint algorithm compared to EI and LEM, both for ordered and pairwise concordance. No algorithm had complete concordance across seizures in any subjects. 

\end{document}